\documentclass{PoS}

\newcommand{\nn}{\nonumber}

\newcommand{\beq}{\begin{equation}}
\newcommand{\eeq}{\end{equation}}
\newcommand{\bea}{\begin{eqnarray}}
\newcommand{\eea}{\end{eqnarray}}
\newcommand{\beqa}{\begin{eqnarray}}
\newcommand{\eeqa}{\end{eqnarray}}

\title{Production of four-jets in single- and double-parton
scattering within high-energy factorization}

\ShortTitle{Four-jet}

\author{Krzysztof Kutak\\       
Institute of Nuclear Physics, Polish Academy of Sciences,\\
{\small\it Radzikowskiego 152, 31-342 Krak\'ow, Poland}\\
        E-mail: \email{Krzysztof.Kutak@ifj.edu.pl}}

\abstract{We report on a first study of 4-jet production in a complete high-energy
factorization (HEF) framework \cite{Kutak:2016mik,Kutak:2016ukc}. We include and discuss contributions
from both single-parton scattering (SPS) and double-parton scattering
(DPS) and compare to the measured data.
The DPS HEF result is
considerably smaller than the one obtained with collinear
factorization. The mechanism leading to this difference is of
kinematical nature. In
contrast to the collinear approach, the HEF approach nicely describes
the distribution of the $\Delta S$ variable, which involves all four
jets and their angular correlations.
}

\FullConference{38th International Conference on High Energy Physics\\
		3-10 August 2016\\
		Chicago, USA}

\begin{document}

\section{Introduction}
So far, complete $(n\geq4)$-jet production via single-parton scattering (SPS) was
discussed only within collinear factorization. Results up to next-to-leading (NLO)
precision can be found in \cite{Bern:2011ep,Badger:2012pf}.
Here we report on recent study of production of four jets including SPS processes and Double-parton scattering (DPS) within high-energy ($k_T$-)factorization (HEF) \cite{Catani:1990xk,Collins:1991ty,Deak:2009xt,Deak:2009ae} \footnote{for overview of the framework see \cite{Sapeta:2015gee}}. 
Double-parton scattering (DPS) was claimed to have been observed for 
the first time at the Tevatron \cite{Abe:1993rv}.
In the LHC era, with much higher collision energies available, 
the field has received a new impulse and 
several experimental and theoretical studies address the problem
of pinning down DPS effects. Even just from purely theoretical point of view, the problem is 
quite subtle \cite{Kotko:2016lej}. As for the non perturbative side, it is in principle necessary, when
considering a double-parton scattering, to take into account 
the correlations between the two partons coming from the same protons 
and involved in the scattering processes. Such an information should be encoded in a set of double parton
distribution functions (DPDFs), generalising usual parton distribution functions (PDFs). Some successful attempts to generalize the usual evolution and to have relevance for phenomenology are becoming 
to appear only recently \cite{Rinaldi:2014ddl,Broniowski:2016trx,Golec-Biernat:2015aza,Golec-Biernat:2016vbt}.

In the meanwhile, phenomenological and experimental studies of double-parton scattering
rely on factorized Ansatz for the DPDFs, which amount
to neglecting  momentum correlations
between partons and introducing an
effective cross section, $\sigma_{eff}$. 
The latter quantity is usually extracted from experimental data.
In the present approach we will use the factorized Ansatz and
concentrate on the difference between leading-order collinear and
high-energy-factorization results. The latter includes effectively
higher-order corrections. 
For most of high-energy reactions the single-parton scattering dominates
over the double-parton scattering. The extraordinary example
is double production of $c \bar c$ pairs \cite{Maciula:2013kd}.
For four-jet production, disentangling the ordinary SPS
contributions from the DPS corrections can be quite challenging for
several reasons: first of all, it is
necessary to define sufficiently sensitive, process-dependent obervables, 
w.r.t. which the DPS differential cross section manifestly dominates at
least in some corners of phase space.

\section{Single-parton scattering production of four jets}%

The HEF factorization formula for the calculation of the inclusive partonic 4-jet cross section at the Born level reads

\bea
\sigma^B_{4-jets} 
&=& 
\sum_{i,j} \int \frac{dx_1}{x_1}\,\frac{dx_2}{x_2}\, d^2 k_{T1} d^2 k_{T2}\,  {\mathcal F}_i(x_1,k_{T1},\mu_F)\, \mathcal{F}_j(x_2,k_{T2},\mu_F) \nn \\
&&
\hspace{-25mm}
\times \frac{1}{2 \hat{s}} \prod_{l=i}^4 \frac{d^3 k_l}{(2\pi)^3 2 E_l} \Theta_{4-jet} \, (2\pi)^4\, \delta\left( x_1P_1 + x_2P_2 + \vec{k}_{T\,1}+ \vec{k}_{T\,2} - \sum_{l=1}^4 k_i \right)\,
\left|{\cal M}(i^*,j^*\rightarrow 4\, part)\right|^2\, . \nn \\
\label{kt_cross}
\eea

Here $\mathcal{F}_i(x_k,k_{Tk},\mu_F)$ is a TMD (Transversal Momentum Dependent)\cite{Angeles-Martinez:2015sea} distribution function for a given type of parton carrying $x_{1,2}$ momentum fractions of the proton and evaluated at the factorization scale $\mu_F$. The index $l$ runs over the four partons in the final state, the
partonic center of mass energy squared is $\hat{s} = 2\,x_1 x_2\, P_i \cdot P_j$; 
the function $\Theta_{4-jet}$ takes into account the kinematic cuts applied and 
$\mathcal{M}\rightarrow 4\, part$ 
is the gauge invariant matrix element for $2\rightarrow 4$ particle scattering with two initial off-shell legs calculated numerically with a numerical package \cite{vanHameren:2016kkz}. It includes symmetrization effects due to identity of particles in the final state and new degrees of freedom introduced via $\vec{k}_{Tk}$, 
which are the parton's transverse momenta, 
i.e. the momenta perpendicular to the collision axis. 
The formula is valid when the $x$'s are not too large and not too small when complications from nonlinearities may eventually arise \cite{Kotko:2015ura}.

\section{Double-parton scattering production of four jets}\label{MPI}
Single-parton scattering contributions are expected to be dominant for high momentum transfer, as it is highly unlikely that
two partons from one proton and two from the other one are energetic enough for two hard scatterings to take place,
as  the behaviour of the PDFs for large $x$ suggests.
However, as the cuts on the transverse momenta of the final state are softened, a window opens
to possibly observe significant double parton scattering effects, as
often stated in the literature on the subject.
First of all, let us recall the formula usually employed for the computation of DPS cross sections, 
adjusting it to the 4-parton final state,
\beq
\frac{d \sigma^{B}_{4-jet,DPS}}{d \xi_1 d \xi_2} = 
\frac{m}{\sigma_{eff}} \sum_{i_1,j_1,k_1,l_1;i_2,j_2,k_2,l_2}
\frac{d \sigma^B(i_1 j_1 \rightarrow k_1 l_1)}{d \xi_1}\, \frac{d \sigma^B(i_2 j_2
\rightarrow k_2 l_2)}{d \xi_2} \, ,
\eeq
where the $\sigma(a b \rightarrow c d)$ cross sections are obtained by
restricting formula (\ref{kt_cross}) to a single
channel and the symmetry factor $m$ is $1$ unless the two hard
scatterings are identical, in which case it is $1/2$, so as to avoid
double counting them. Above $\xi_1$ and $\xi_2$ stand for generic
kinematical variables for the first and second scattering, respectively.
The effective cross section $\sigma_{eff}$ can be interpreted 
as a measure of the transverse correlation of the two partons inside 
the hadrons, whereas the possible longitudinal correlations are usually
neglected.
In this paper we use $\sigma_{eff}$ provided by the CDF, D0
collaborations and recently confirmed by the LHCb collaboration
$\sigma_{eff}$ = 15 mb, when all SPS mechanisms of double charm production are included.
%

\subsection{Comparison to CMS data}

As discussed in Ref.~\cite{Maciula:2015vza}, so far the only experimental analysis 
of four-jet production relevant for the DPS studies was realized
by the CMS collaboration \cite{Chatrchyan:2013qza}. 
The cuts used in this analysis are $p_T> 50$ GeV for the first and
second jets, $p_T > 20$ GeV for the third and fourth jets,
$|\eta| < 4.7$ and the jet cone radius parameter $\Delta R > 0.5$. 

It goes without saying that the LO result with soft cuts applied needs refinements from NLO contributions.
For this reason, in the following we will always perform comparisons only to 
data (re)normalised to the total (SPS+DPS) cross sections.
What is interesting in the HEF result, 
compared to collinear factorization, is the dramatic damping of the DPS
contribution.
%
\begin{figure}[h]
\begin{center}
\begin{minipage}{0.47\textwidth}
 \centerline{\includegraphics[width=1.0\textwidth]{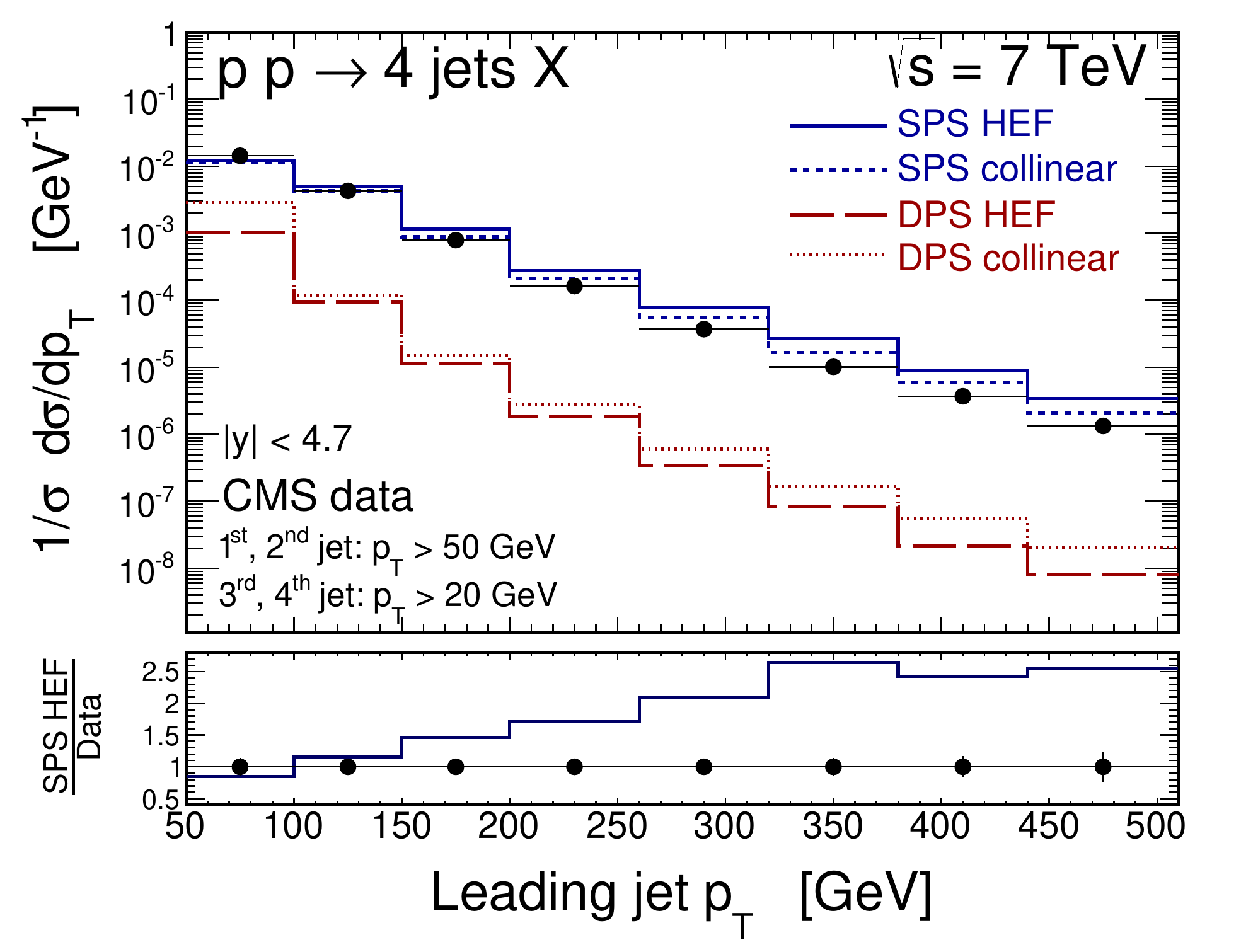}}
\end{minipage}
\hspace{0.5cm}
\begin{minipage}{0.47\textwidth}
 \centerline{\includegraphics[width=1.0\textwidth]{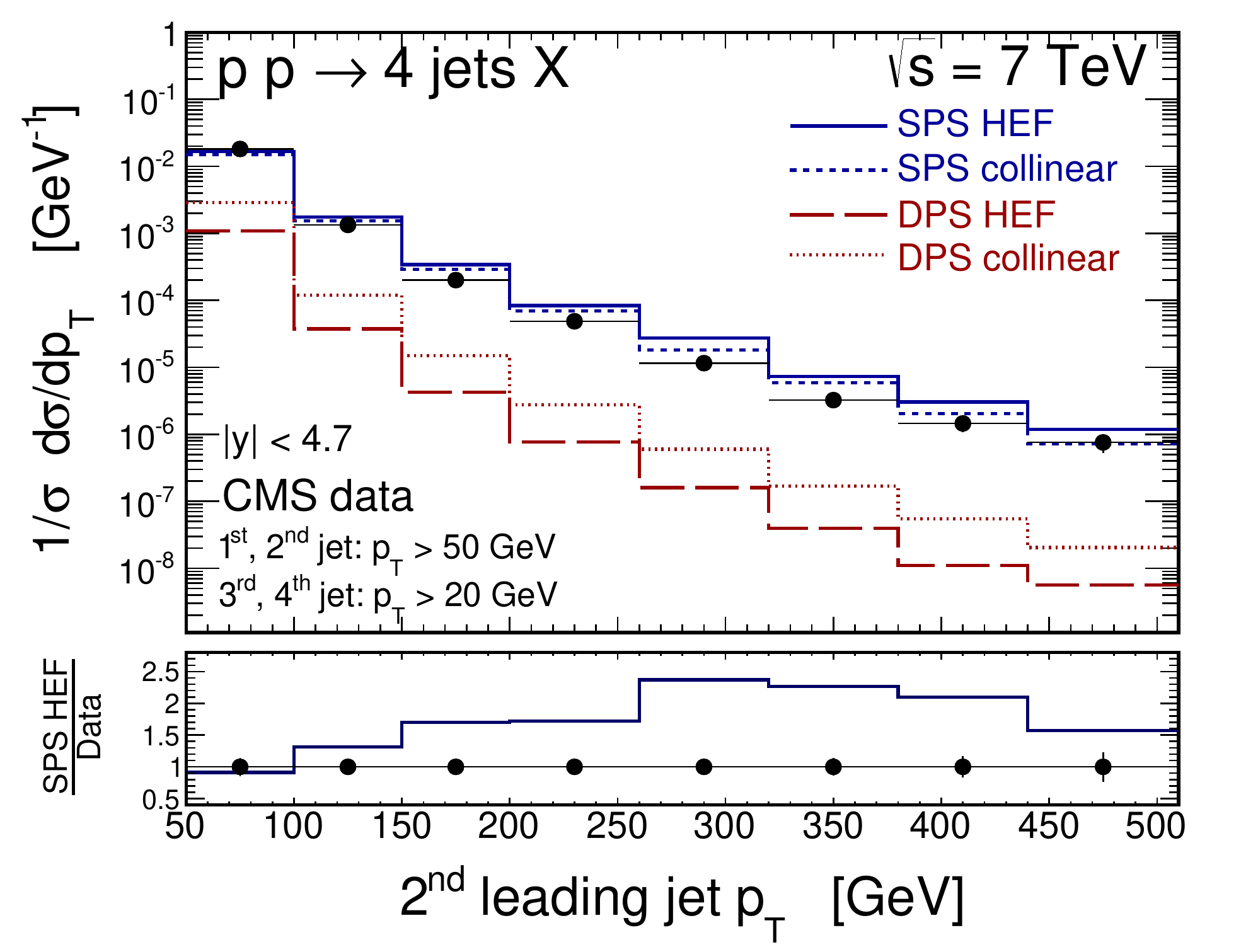}}
\end{minipage}
\end{center}
\caption{
Comparison of the LO collinear and HEF predictions to
the CMS data for the 1st and 2nd leading jets. In addition we show the ratio of the SPS HEF result to the CMS data.}
\label{CMS_pT_12}
\end{figure}
%
\begin{figure}[h]
\begin{center}
\begin{minipage}{0.47\textwidth}
 \centerline{\includegraphics[width=1.0\textwidth]{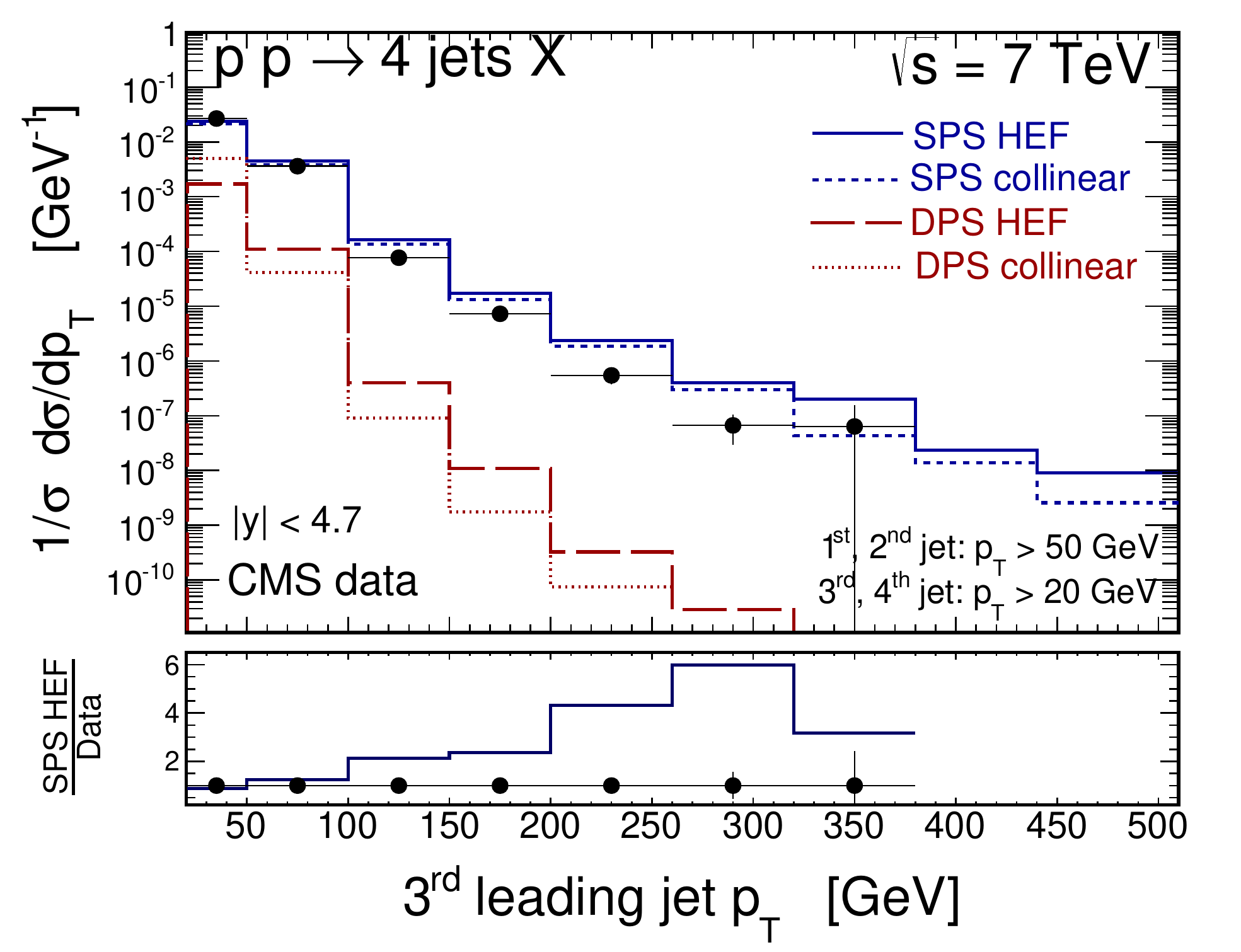}}
\end{minipage}
\hspace{0.5cm}
\begin{minipage}{0.47\textwidth}
 \centerline{\includegraphics[width=1.0\textwidth]{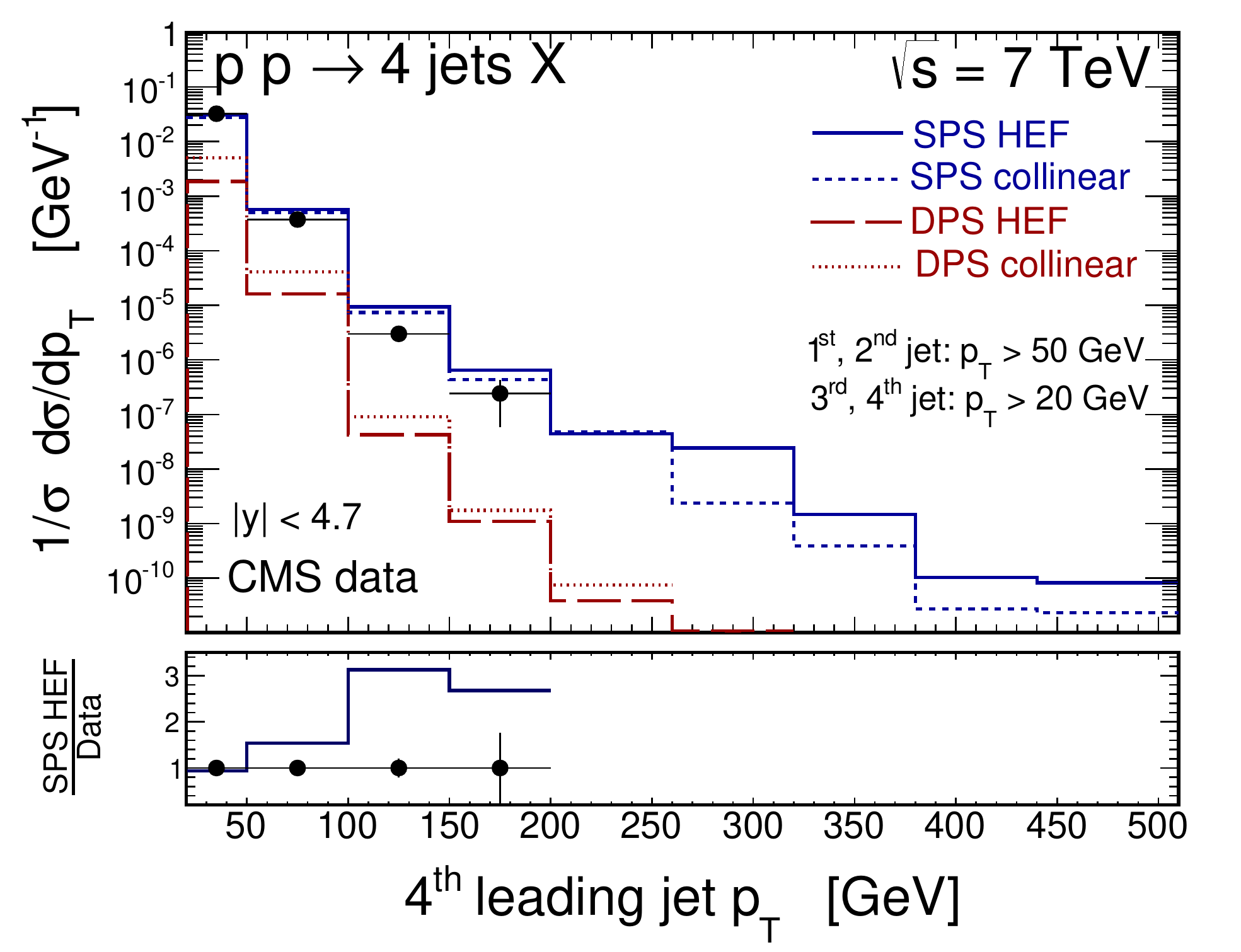}}
\end{minipage}
\end{center}
\caption{
Comparison of the LO collinear and HEF predictions to 
the CMS data for the 3rd and 4th leading jets. In addition we show the ratio of the SPS HEF result to the CMS data.}
\label{CMS_pT_34}
\end{figure}
%
In Figs.~\ref{CMS_pT_12} and \ref{CMS_pT_34} we compare the predictions
in HEF to the CMS data. Here both the SPS and DPS
contributions are normalized to the total cross section,
i.e. the sum of the SPS and DPS contributions.
In all cases the renormalized transverse momentum distributions agree
with the CMS data. However, the absolute cross sections obtained in this case
within the HEF approach are too large.

Not only transverse momentum dependence is interesting.
The CMS collaboration extracted also a more complicated observables
\cite{Chatrchyan:2013qza}.
One of them, which involves all four jets in the final state,
is the $\Delta S$ variable, defined in Ref.~\cite{Chatrchyan:2013qza} as the angle between pairs of
the harder and the softer jets,
%
%
where $\vec{p}_T(j_i,j_k)$ stands for the sum of the transverse momenta of the two jets in arguments.

\begin{figure}[h]
\begin{center}
\begin{minipage}{0.47\textwidth}
 \centerline{\includegraphics[width=1.0\textwidth]{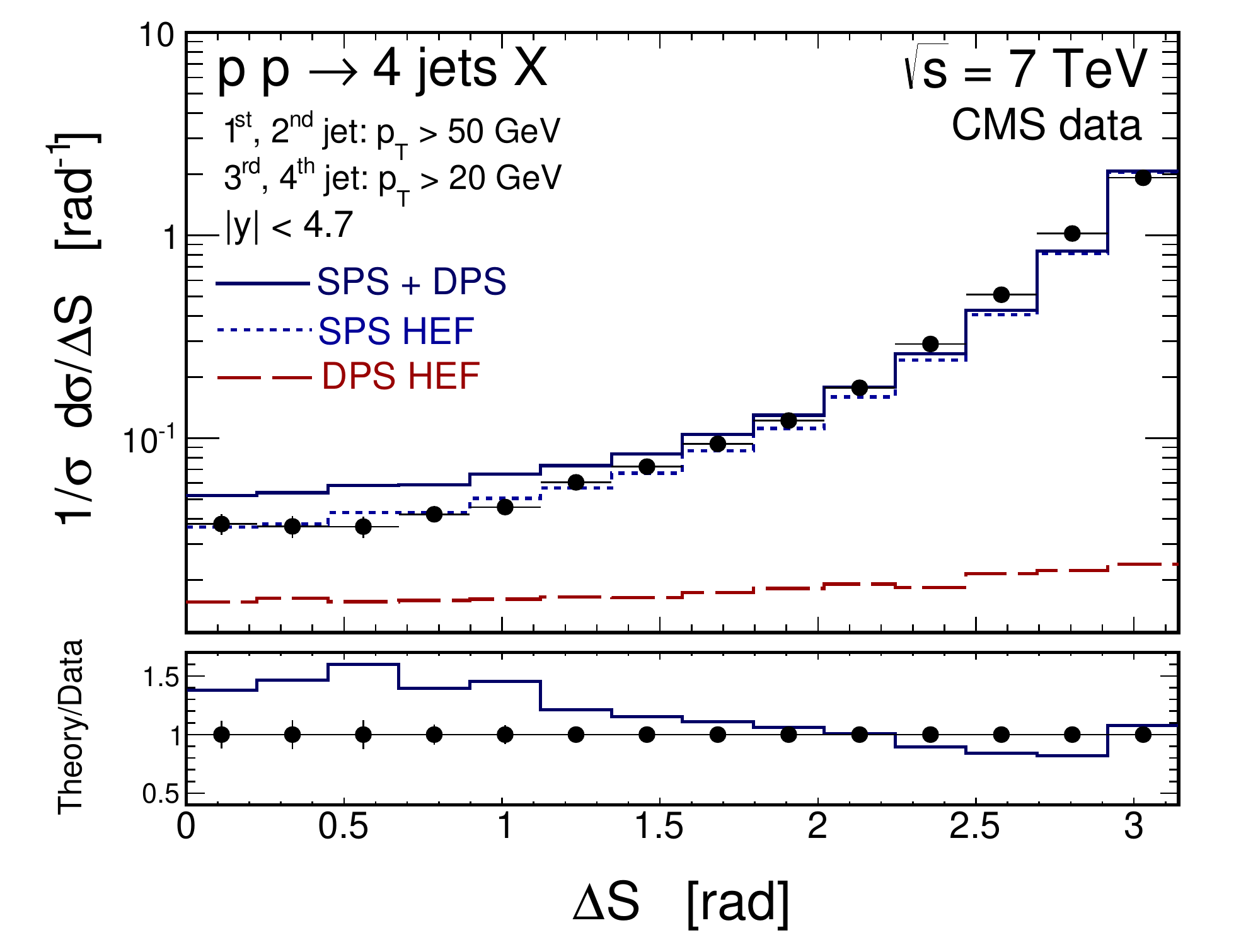}}
\end{minipage}
\end{center}
\caption{
Comparison of the HEF predictions to the CMS data for $\Delta S$ spectrum. In addition we show the ratio of the (SPS+DPS) HEF result to the CMS data.}
\label{fig:CMS_DS}
\end{figure}

In Fig.~ \ref{fig:CMS_DS} we present our HEF prediction for
the normalized to unity distribution in the $\Delta S$ variable. 
Our HEF result approximately agrees with the
experimental $\Delta S$ distribution. 
In contrast the LO collinear approach leads to 
$\Delta S$ = 0, i.e. a Dirac-delta peak at $\Delta S$ = 0 for 
the distribution in $\Delta S$.
For the DPS case this is rather trivial. The two hard and two soft 
jets come in this case from the same scatterings and are back-to-back
(LO), so each term in the argument of $arccos$ is zero 
(jets are balanced in transverse momenta).
For the SPS case the transverse momenta of the two jet pairs 
(with hard jets and soft jets) are identical and have opposite direction
(the total transverse momentum of all four jets must be zero from the momentum conservation).
Then it is easy to see that the argument of $arccos$ is just -1.
This means $\Delta S$ = 0.
The above relations are not fullfilled in the HEF
approach. The SPS contribution clearly dominates and approximately gives the shape
of the $\Delta S$ distribution.
It is anyway interesting that we describe the data
via pQCD effects within our HEF approach which 
are in Ref.~\cite{Chatrchyan:2013qza} described by parton-showers and soft MPIs. 

%
%

\section*{Acknowledgments}
The work of K.K. have been supported by Narodowe Centrum Nauki
with Sonata Bis grant DEC-2013/10/E/ST2/00656. The paper is based on common publication with Rafal Maciula, Mirko Serino,
Andreas van Hameren, Antoni Szczurek.

\end{document}